\documentclass[a4paper,11pt]{article}
\pdfoutput=1 

\usepackage{jinstpub} 

\usepackage{lineno}

\newcommand{\gevc}{GeV/\textit{c}}

\newcommand{\pt}{$p_{\mathrm{T}}$}

\newcommand{\dedx}{$dE/dx$ }
\newcommand{\el}{$e^{-}$ }

\title{Calibration and Performance of the Upgraded ALICE Inner Tracking System}

\author[a]{Andrea Sofia Triolo}

\affiliation[a]{CERN,\\Geneva, Switzerland}

\emailAdd{andrea.sofia.triolo@cern.ch}

\abstract{The ALICE Experiment at the Large Hadron Collider (LHC) underwent a major upgrade during the Long Shutdown 2. Several subsystems have been improved, including the ALICE Inner Tracking System (ITS), which has been entirely replaced. The new pixel-only tracker (ITS2) consists of 7 layers of Monolithic Active Pixel Sensors (MAPS) featuring a pixel size of 27×29~$\upmu$m², with an intrinsic spatial resolution of 5~$\upmu$m. With 24120 sensors and 12.5 billion pixels, this detector covers an active area of about 10 m$^2$ and represents the largest application of the MAPS technology in a high-energy physics experiment to date. The most significant improvements introduced by the ITS2 to the ALICE experiment include a reduction in the impact parameter resolution to approximately 30 $\upmu$m in both the r$\varphi$ and z coordinates at a transverse momentum of 1 GeV/c. This is a factor of 3 improvement over the previous detector. Additionally, the standard ITS readout rate has been increased from 1 kHz to 67 kHz in Pb-Pb collisions and to 202 kHz in proton-proton collisions. To ensure stable operations and maintain high data quality a regular calibration is performed, which consists in establishing the charge threshold and the noisy channels of the detector.  The ITS2 has been successfully commissioned for LHC Run3, and already operated during proton-proton and Pb-Pb collisions at LHC with excellent performance. This contribution gives an overview of the operational procedures required to maintain an optimal data quality, along with results obtained from calibration and the performance achieved during the LHC Run 3.}


\collaboration[c]{on behalf of the ALICE collaboration}

\proceeding{25$^{\text{th}}$ International Workshop on Radiation Imaging Detectors\\
  30$^{\text{th}}$ June - 04$^{\text{th}}$ July 2024\\
  Lisbon, Portugal}

\begin{document}
\maketitle
\flushbottom

\section{The Upgraded ALICE Inner Tracking System}\label{Alice_upgrade}
ALICE (A Large Ion Collider Experiment) \cite{ALICE} is one of the four major experiments at the CERN Large Hadron Collider (LHC). It was designed to investigate the properties of the strongly interacting matter, created in the proton-proton and Pb--Pb collisions at the LHC. To fulfill the requirements of the LHC Run 3 physics program \cite{TDR}, the experiment underwent an upgrade that was completed in Spring 2021. An important part of the upgrade concerned the ALICE Inner Tracking System (ITS), the innermost ALICE detector, dedicated to localizing primary and secondary vertices and reconstructing particle tracks. The previous ITS (ITS1) has been completely replaced with a new detector (ITS2) to improve the tracking efficiency, the \pt \ resolution, the impact parameter resolution, and the readout capabilities. A key feature of the upgrade was the replacement of the beam pipe at the center of ALICE with a smaller one, passing from a radius of 28 mm to a radius of 18 mm. This allowed the first detection layer of the ITS to be closer to the interaction point, significantly increasing the pointing resolution. Another key feature of the upgrade was the addition of a detection layer, passing from 6 layers for the ITS1 to 7 layers for the ITS2, contributing to an enhancement of the tracking efficiency and \pt \ resolution at low \pt.
\\
The ITS2 is now composed of 7 cylindrical and concentric layers of Monolithic Active Pixel Sensors (MAPS) named ALPIDE (ALice Pixel Detector) \cite{ALICE_upgrade}, whose features will be discussed in Section \ref{alpide}. The first layer is located 2.2 cm away from the center of the beam pipe, while the last layer is located at a radius of 39 cm. The 7 layers are divided into Inner barrel (IB, layers from 0 to 2), and Outer Barrel (OB, layers from 3 to 6). Moreover, each layer is azimuthally segmented into mechanically independent units called staves, for a total of 192 staves composing the detector. Each stave is composed of ALPIDE MAPS, for a total of 24120 chips in the entire detector and a covered active area of 10 m$^2$. These features make the ITS2 the largest pixel-based detector in High-Energy Physics to date. Thanks to innovative technology used to build the layers, the material budget of the ITS2 has been reduced to 0.36\% $X_0$/layer for the IB and 1.1\% $X_0$/layer for the OB. Having a low material budget is of fundamental importance to improve the tracking performance of the detector, especially for low-momentum particles whose resolution is mostly influenced by multiple scattering in the material.

\subsection{ALPIDE MAPS}\label{alpide}
\begin{figure}
    \centering
    \includegraphics[width=0.5\linewidth]{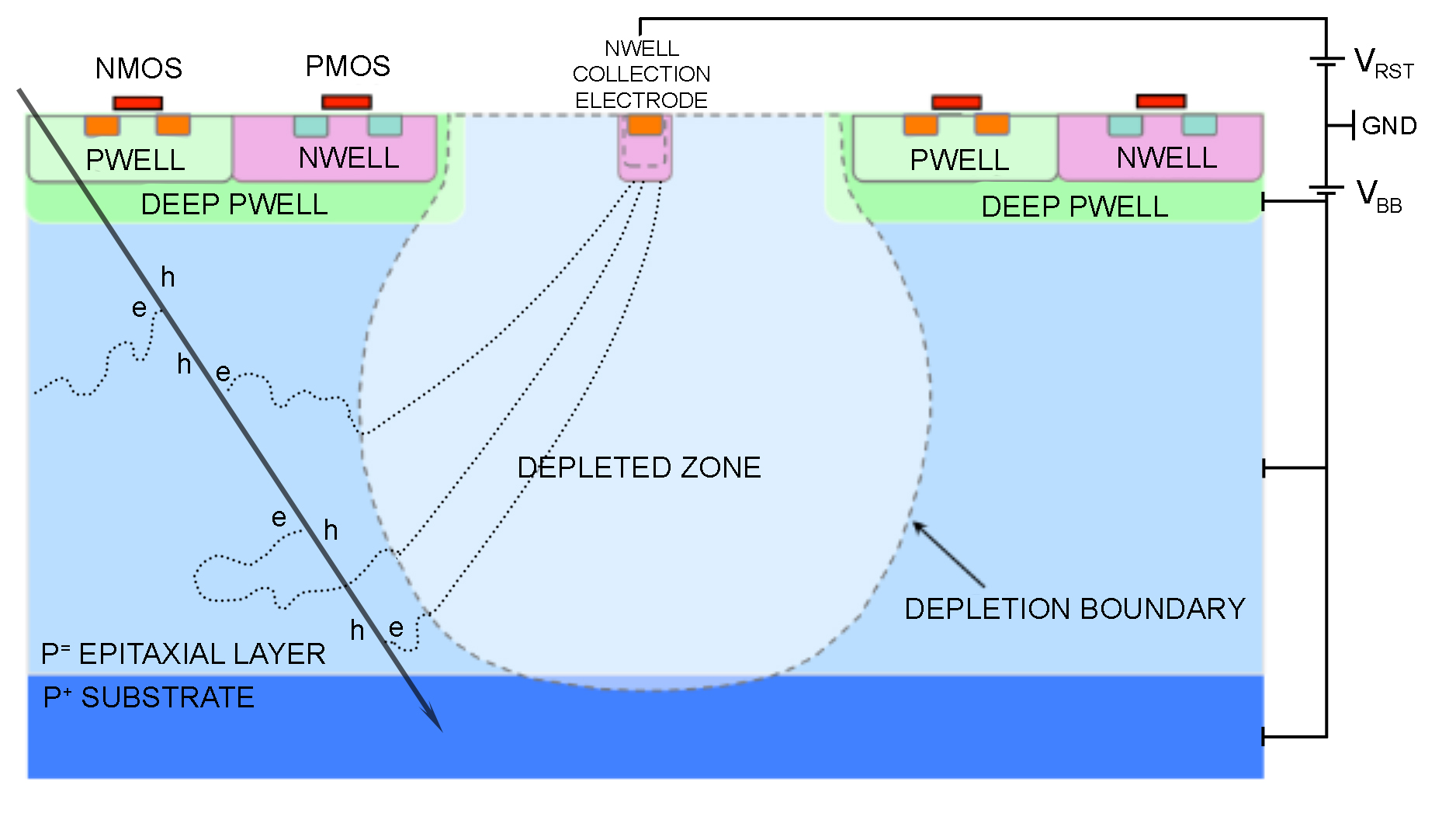}
    \caption{Cross section of the ALPIDE pixel cell.}
    \label{fig:crossALPIDE}
\end{figure}

The ALPIDE MAPS \cite{ALICE_upgrade} has been developed for the ALICE upgrade using the TowerJazz 180 nm CMOS Imaging Sensor process. Each chip measures 1.5 cm $\times$ 3 cm, and includes a matrix of 512 pixel rows in the $r\phi$ direction and 1024 pixel columns in the $z$ direction, with every pixel cell having a pitch of 29.24 $\times$ 26.88 $\upmu$m$^2$. 
The signal-sensing elements in the chip are n-well diodes, which have an area 100 times smaller than the area of the pixel cell. A deep p-well implant allows the implementation of full CMOS circuitry inside the pixel matrix as it shields the epitaxial layer from the n-wells of the PMOS transistors. Without this feature, the charge collection on the sensing diodes would be compromised by the n-wells of the PMOS transistor. A schematic view of the cross section of the ALPIDE pixel cell is shown in Figure \ref{fig:crossALPIDE}.
\\ The pixel signal is amplified and discriminated at the pixel level, leading to a binary output. Each pixel cell contains a sensing diode, a pulse injector capacitor to inject test changes during calibration, a front-end circuit that amplifies, shapes and discriminates the signal, and a digital section that includes 3 hit storage registers, a pixel masking register and the pulsing logic. The address of the fired pixel is sent to the \textit{chip periphery}, a peripheral region of the chip measuring 1.2~$\times$~30 mm$^2$ where the analog biasing, control, readout and interfacing functionalities are implemented. The chip periphery contains fourteen 8-bit analog DACs (Digital-to-Analog Converters) for biasing the pixel front-ends. Test beam results \cite{ALICE_upgrade} have shown that the sensor has an intrinsic spatial resolution of 5 $\upmu$m and a detection efficiency above 99\%. About 60 $e^-$/$\upmu$m are generated in the sensor by a traversing Minumum-Ionizing Particle (MIP), therefore about 1500 \el are generated for a MIP normal incidence in the 25 $\upmu$m epitaxial layer of the ALPIDE sensors, shared between adjacent pixels.

\subsection{ITS2 Data Readout}
Each detector stave is connected to an FPGA-based Readout Unit (RU) providing control, power, trigger and monitoring. The 192 RUs are then connected to a set of 22 Common Readout Units (CRUs) installed in 13 First Level Processors (FLPs). Data read from the RUs are aggregated in the CRUs and sent to the Event Processing Nodes (EPNs), composed by 340 servers where data are stored, reconstructed, and processed online or offline. Here Quality Control (QC) tasks are available to synchronously monitor data in output from the detector. The data readout and the subsequent processing are managed by the O2 framework \cite{O2}, the ALICE computing system for the LHC Run 3. The standard read-out mode for the detector is the continuous mode: the RUs internally generate periodic triggers to the sensors with a minimal gap between the strobes. This allows a continued read-out of the sensors, whose output data are then segmented in time frames of fixed duration.

\section{ITS and ALICE performance during LHC Run 3}
\begin{figure}
    \centering
    \includegraphics[width=0.4\linewidth]{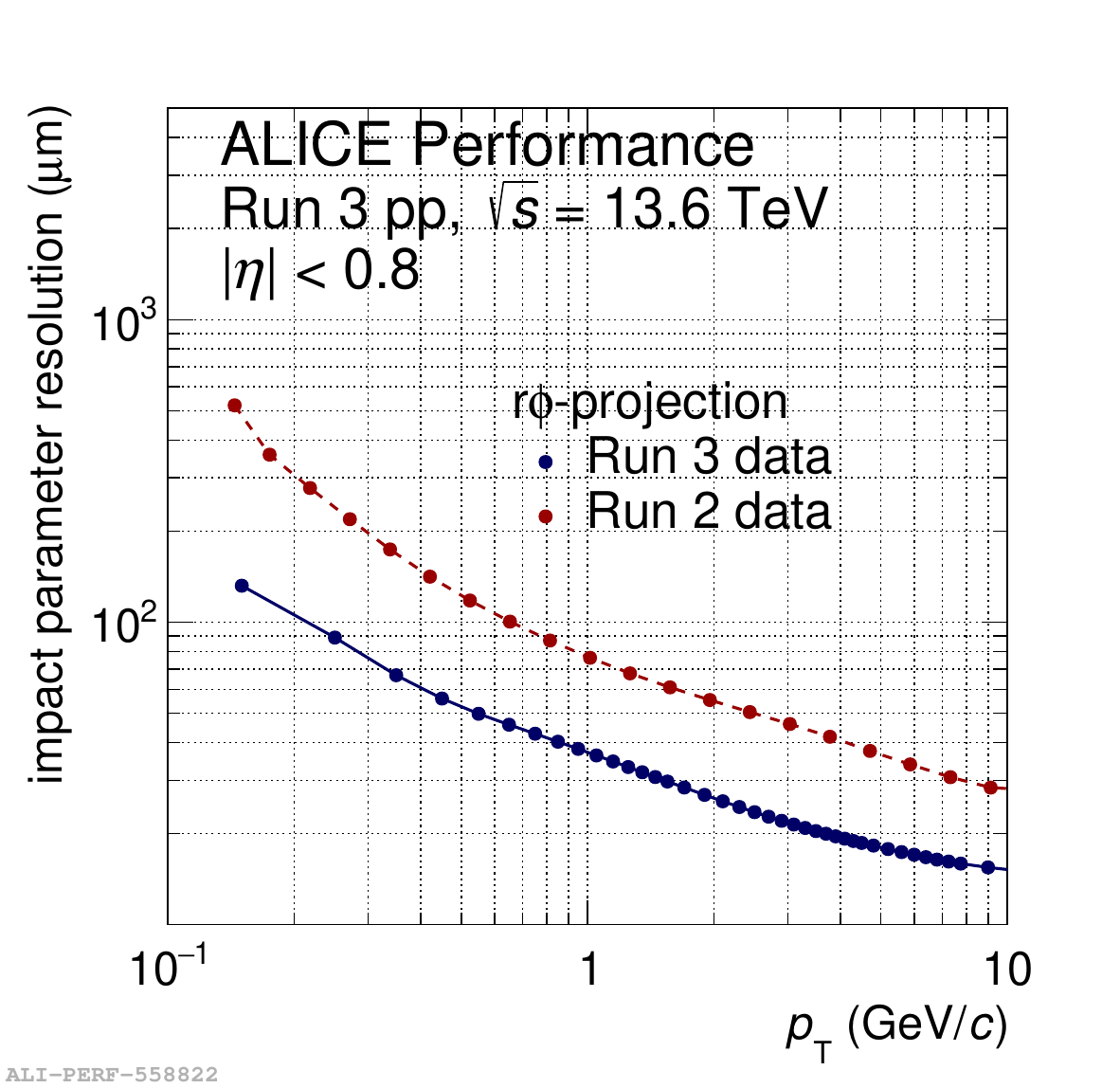}
    \includegraphics[width=0.4\linewidth]{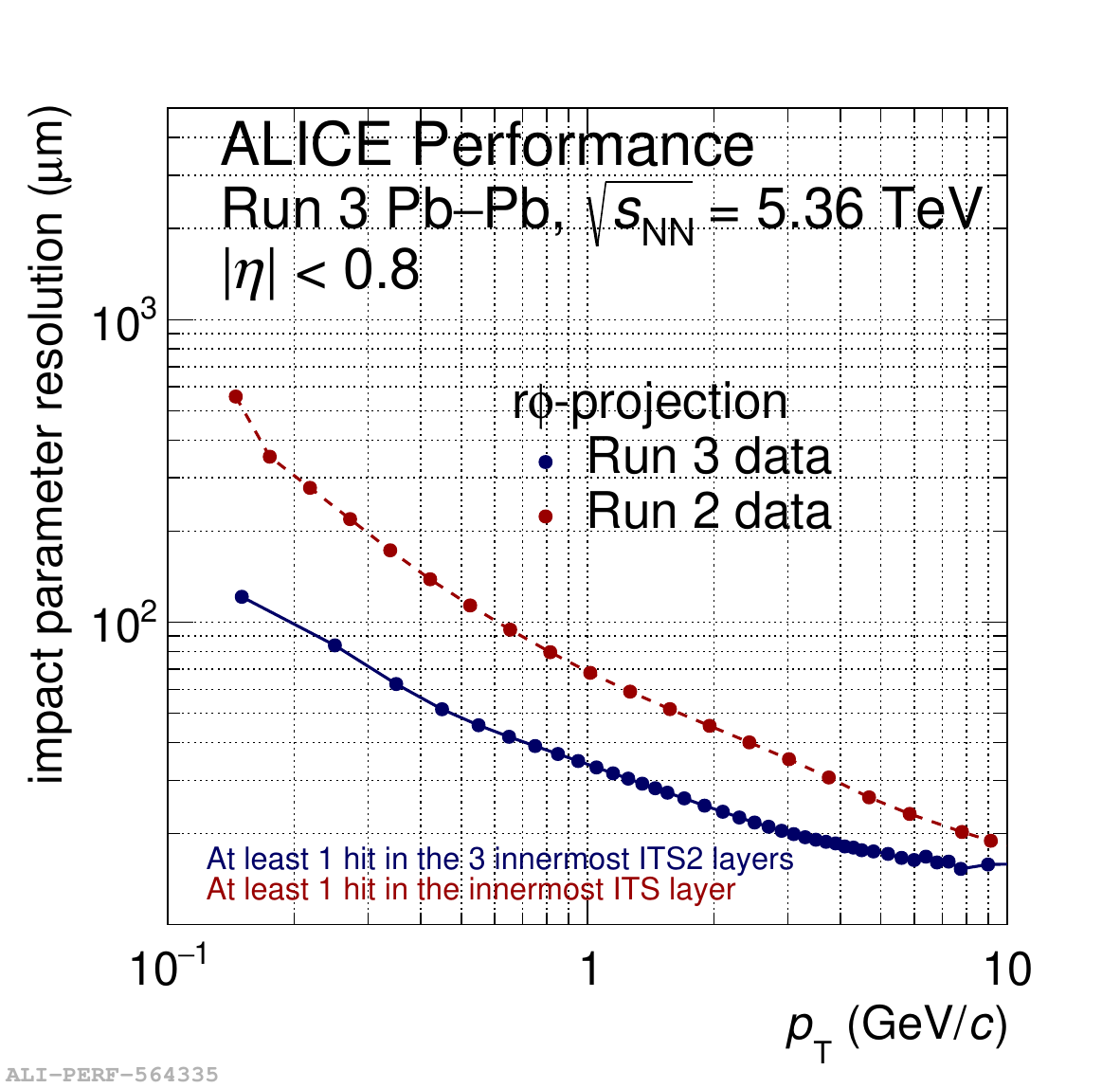}
    \caption{Impact parameter resolution evaluated at the primary vertex in $r\phi$ vs \pt in pp collisions (left) and Pb--Pb collisions (right) in Run 3 compared with the same quantity measured in the corresponding collisions in Run 2.}
    \label{fig:impactParameter}
\end{figure}

\begin{figure}
    \centering
    \includegraphics[width=0.4\linewidth]{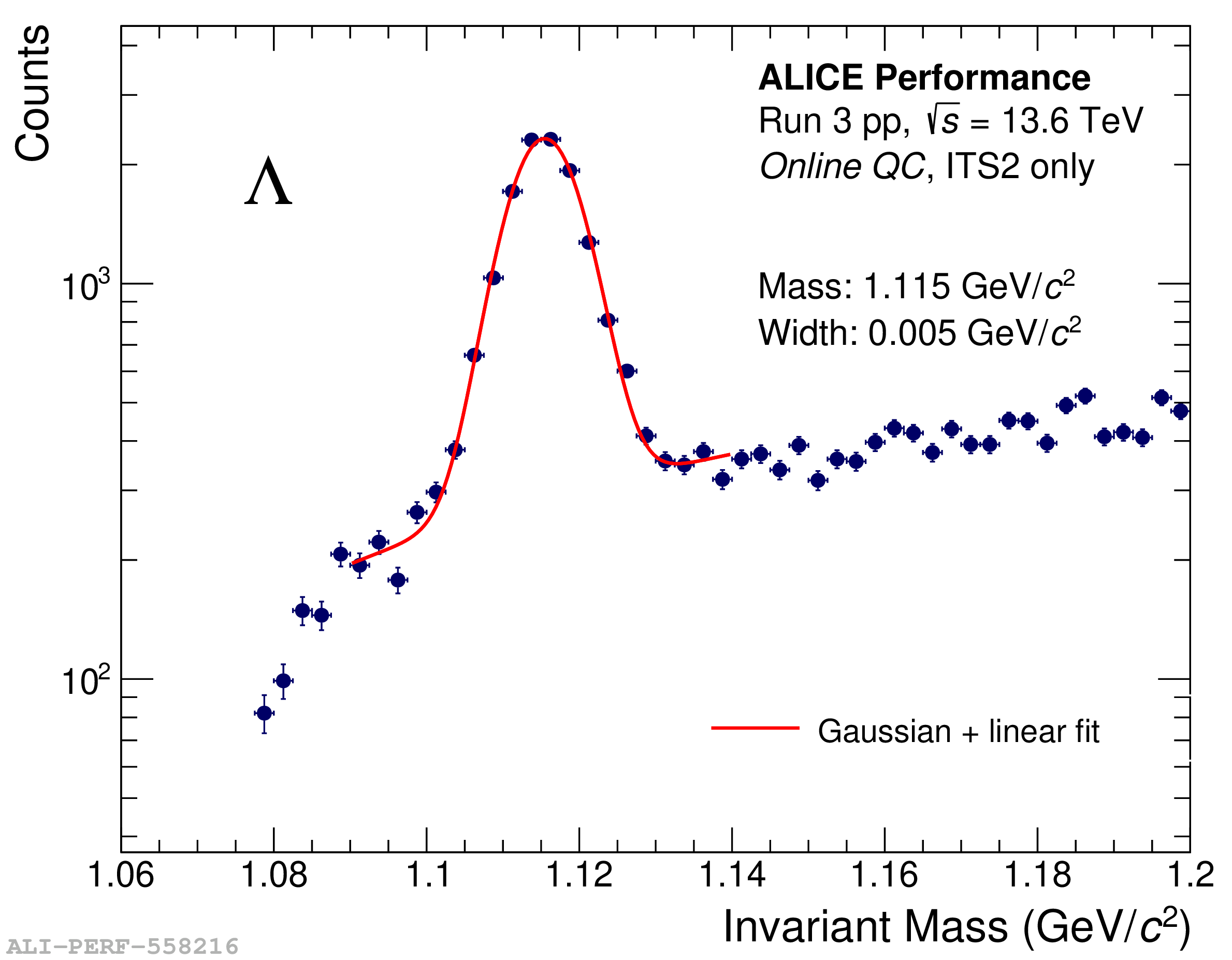}
    \includegraphics[width=0.4\linewidth]{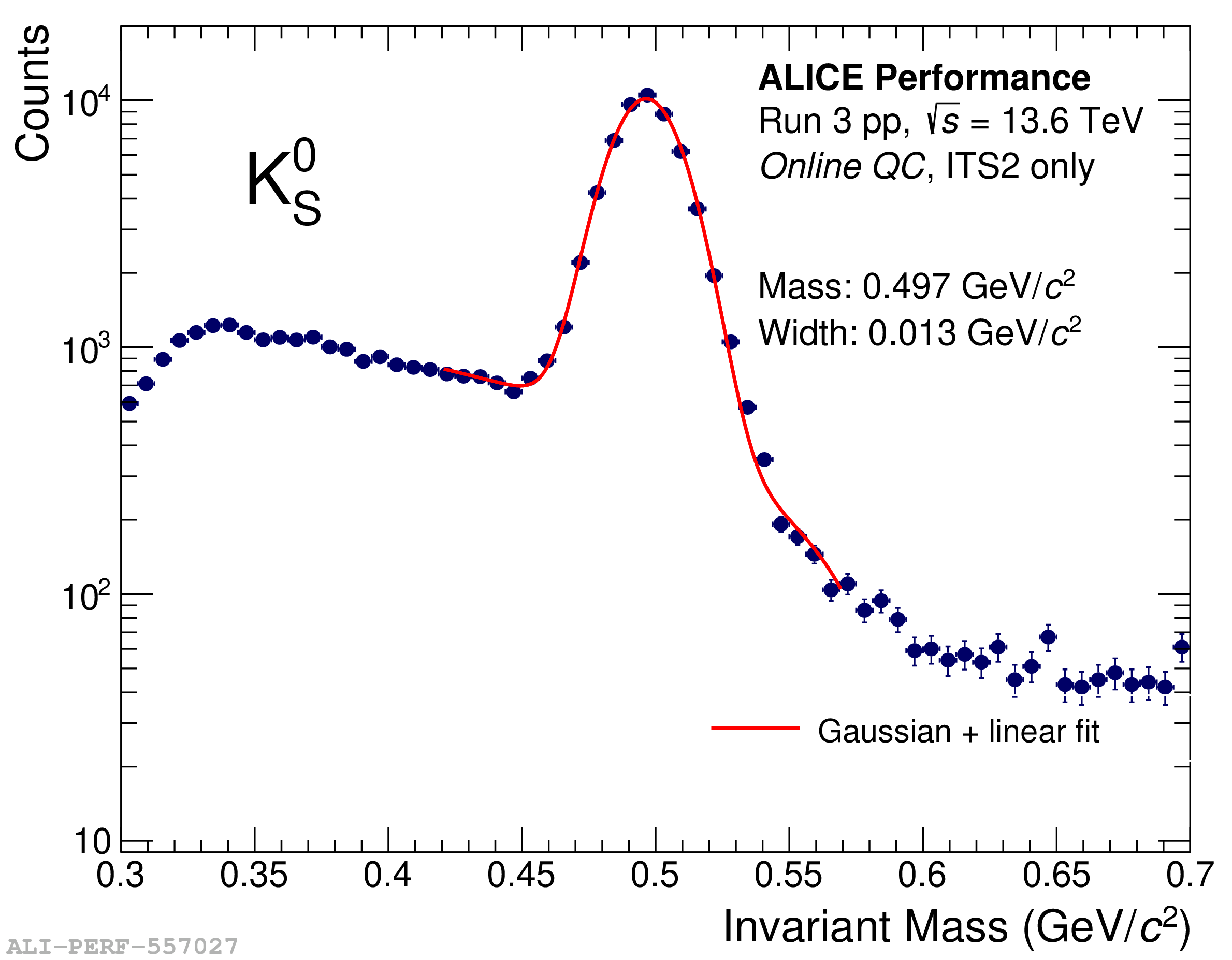}
    \caption{Invariant-mass peaks of $\Lambda$ (left) and $K^0_s$ (right) obtained with ITS standalone tracks synchronously reconstructed in pp and Pb--Pb collisions during LHC Run 3.}
    \label{fig:lambdaandkaon}
\end{figure}

The recorded luminosity integrated so far during Run 3 reached about 38 pb$^{-1}$ in pp collisions at $\sqrt{s}$=13.6 TeV and about 2 nb$^{-1}$ in Pb--Pb collisions at $\sqrt{s_{NN}}$= 5.36 TeV. The ITS2 has been successfully tested up to 4 MHz of pp interaction rate, even if the nominal interaction rate is set to 500 kHz for pp collisions and 45 kHz for Pb--Pb collisions.
The default framing rates for the ITS2 are of 202 kHz during proton collisions and 67 kHz during ion collisions.
\\
\\
As shown in Figure \ref{fig:impactParameter}, an improvement of factor 2 has been measured for the impact parameter resolution at \pt \ = 1 \gevc \ during Run 3 with respect to Run 2, both for pp and Pb--Pb collisions.
Figure \ref{fig:lambdaandkaon} shows an example of the invariant-mass peak of $\Lambda$ and $\mathrm{K}^0_\mathrm{s}$ obtained from the QC task, using only ITS standalone tracks reconstructed by matching positive and negative particle tracks in the ITS layers. These plots demonstrate the excellent performance of the ITS2 tracking. 
Another important performance achievement that can be obtained in Run 3 derives directly from the reduced distance between the primary vertex and the innermost layers of the ITS2, which range from 2.2 cm to 4 cm. This allows the tracking of charged weak-decaying particles before their decay, via the \textit{strangeness tracking} algorithm \cite{sstracking}. This technique exploits the proximity between the first ITS2 layer and the primary vertex to directly measure the hits of the weakly decaying hadron before its decay, and combine this information with the one from the decay daughters, highly enhancing the pointing resolution. This technique opens the possibility to study non-prompt cascades, hypernuclei, and exotic bound states with an unprecedented resolution.

\section{ITS2 Calibration}
The calibration of the ITS2 is an important step required to ensure stable operation and high data quality. Since the high number of pixels of the detector, on the order of 12.5 $\times$ 10$^9$, the calibration of the ITS2 constitutes an unprecedented challenge.
\\
A general standard calibration procedure involves the tuning of the pixel thresholds and the masking of the noisy pixels. The ITS2 configuration and calibration are entirely managed by the Detector Control System (DCS), and data read-out and subsequent processing are integrated into the ALICE O2.

\subsection{Threshold calibration}
\begin{figure}
    \centering
    \includegraphics[width=0.7\linewidth]{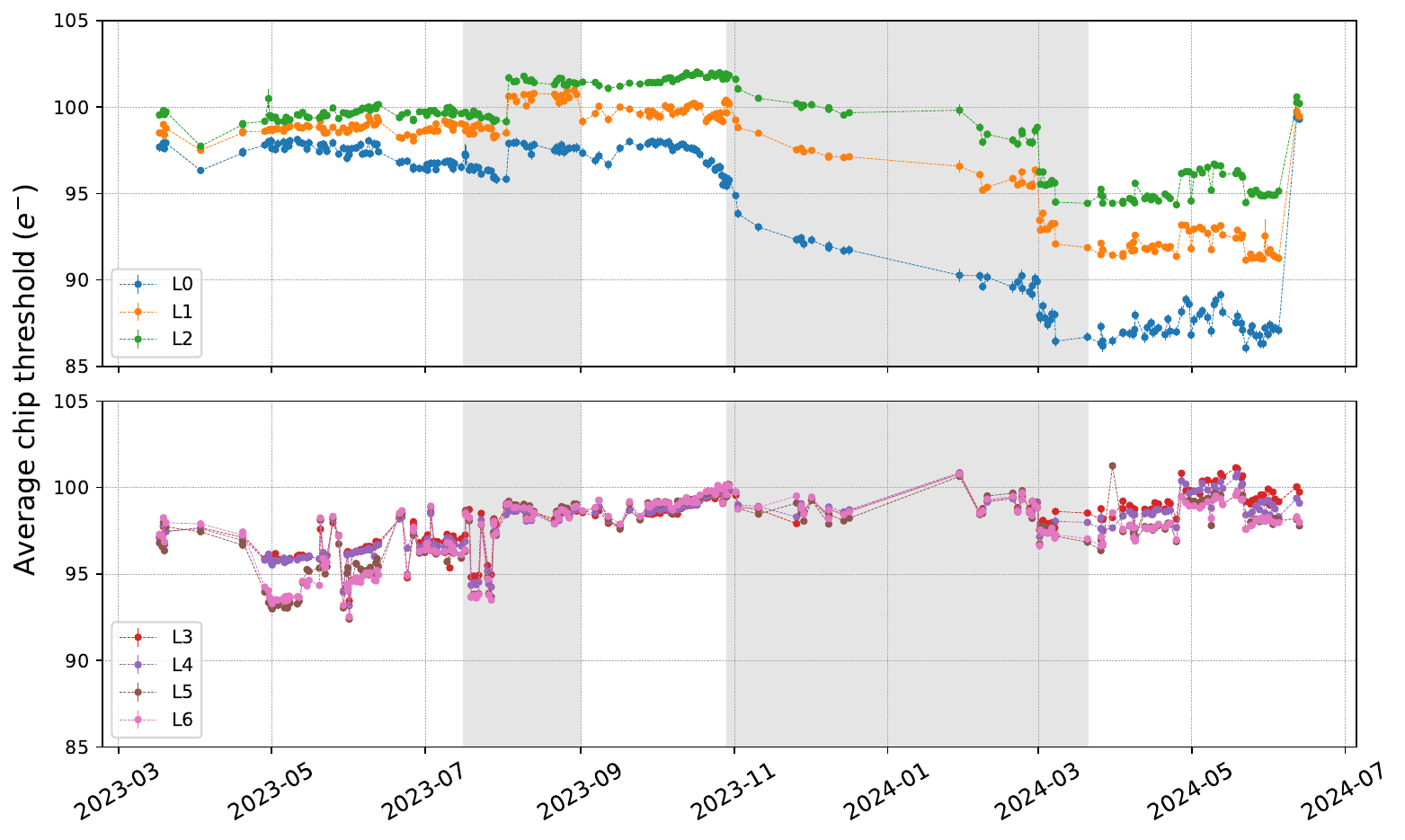}
    \caption{Evolution of the average threshold per layer, from March 2023 to July 2024. The gray background color indicates the long periods without beam during and after the 2023 data-taking phase.}
    \label{fig:thrtrend}
\end{figure}

The threshold calibration involves a threshold tuning, whose purpose is to set the operating point of the detector, and the threshold scan, with which is possible to measure the threshold of the chips.
\\
The \textit{threshold tuning }is performed by setting the values of the \texttt{VCASN} and \texttt{ITHR} DACs, which influence respectively the baseline and the shape of the analogue response of the ALPIDE chips, and therefore change the set threshold. More in detail, an increase of \texttt{VCASN} produces an exponential decrease of the threshold, while an increase of \texttt{ITHR} produces a linear increase of the threshold. \texttt{VCASN} and \texttt{ITHR} values are set per chip, therefore what it is actually tunable is the average threshold of a chip. The target value chosen for the threshold is 100 $e^{-}$, resulting from a compromise between having a good detection efficiency and a low fake-hit rate \cite{ALICE_upgrade}. The tuning of the DACs is done by injecting an analog pulse with the fixed amplitude of 100 \el into the input of the pixels and counting the number of hits registered in the pixels. Then the DAC setting is changed and the injection of the fixed charge and the hit counting are repeated. This is repeated 50 times for each DAC setting. The pixel response vs. the DAC setting is fitted with an error function to extract the 50\% point, which represents the DAC value for which the threshold set is 100 $e^{-}$. Based on the experience acquired during commissioning, a small percentage of pixels is enough to obtain the complete information for the whole detector, hence the threshold tuning is usually done on only about 1\% of pixels equally distributed on each chip.
\\ \\
After the threshold tuning, the value of the threshold that has been set is measured via a \textit{threshold scan}. The threshold can be measured pixel by pixel, but the meaningful information is the average threshold per chip since the tuning is performed to set an average threshold per chip at the desired value. The scan is performed by injecting analog pulses ranging from 0 to 500 electrons into the pixels, keeping the \texttt{VCASN} and \texttt{ITHR} values fixed at the value set during the tuning. As for the threshold tuning, the response of the chip is fitted. From the fit, it is possible to extract the 50\% point that represents the threshold and the $\sigma$, which represents the ENC noise. Also for the threshold scan it is sufficient to perform it only on around 2\% of pixels, equally distributed in the chip matrix. 
\\
Results from the threshold scan are analyzed not only to assess the quality of the tuning, but threshold scans are also performed daily and the results are stored to monitor the stability of the threshold of the detector over time, as it is reported in Figure \ref{fig:thrtrend}.
In particular, it is possible to observe that the threshold has decreased over the months. This was due to radiation effects, known from the R\&D phase, consistent with the fact that the largest decrease has been seen from the innermost layers. A new tuning was performed in June 2024, having as a result the compensation for the decrease of the threshold accumulated over months. This plot proves that continuously monitoring the threshold is important to spot high changes in the threshold and correct them to ensure stable data taking.

\subsection{Noise calibration}
\begin{figure}
    \centering
    \includegraphics[width=0.8\linewidth]{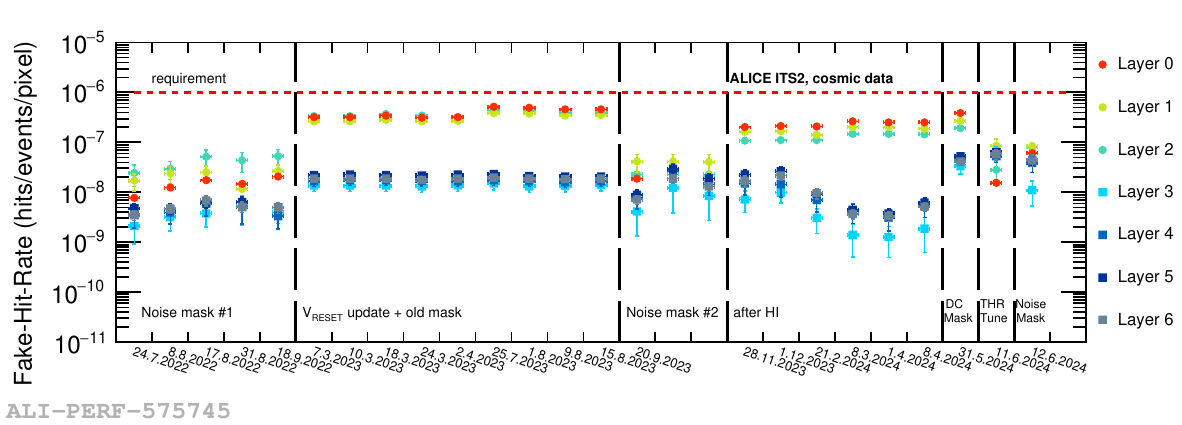}
    \caption{Trend of the fake-hit rate averaged for every detector layer after the application of the noise masks.}
    \label{fig:fhr}
\end{figure}

The noise calibration is performed to identify and mask the noisy pixels of the detector. The noise calibration consists of a standard physics run without the presence of beam or charge injections, so-called \textit{cosmic run}. Pixels exceeding 10$^{-2}$ hits/event/pixel for the IB and 10$^{-6}$ hits/event/pixel for the OB are tagged as noisy and masked. The choice to set different thresholds for IB and OB was made to avoid efficiency losses in the IB, and because the number of hits registered in OB would be dominated by noise if such a less strict cut would be applied, considering the lower occupancy of the OB layers. With these requirements, the number of noisy pixels masked is 0.004\% of the 12.5 billion pixels in the entire detector.
Figure \ref{fig:fhr} shows the average FHR per layer during different periods, after the application of different noise masks.
Even if the overall FHR value slightly changes over time, it always remains below the 10$^{-6}$ hits/event/pixel, which is the maximum level imposed by design to achieve a good track reconstruction performance \cite{TDR}.

\section{Proof of concept for Particle Identification with MAPS}
\begin{figure}
    \centering
    \includegraphics[width=0.7\linewidth]{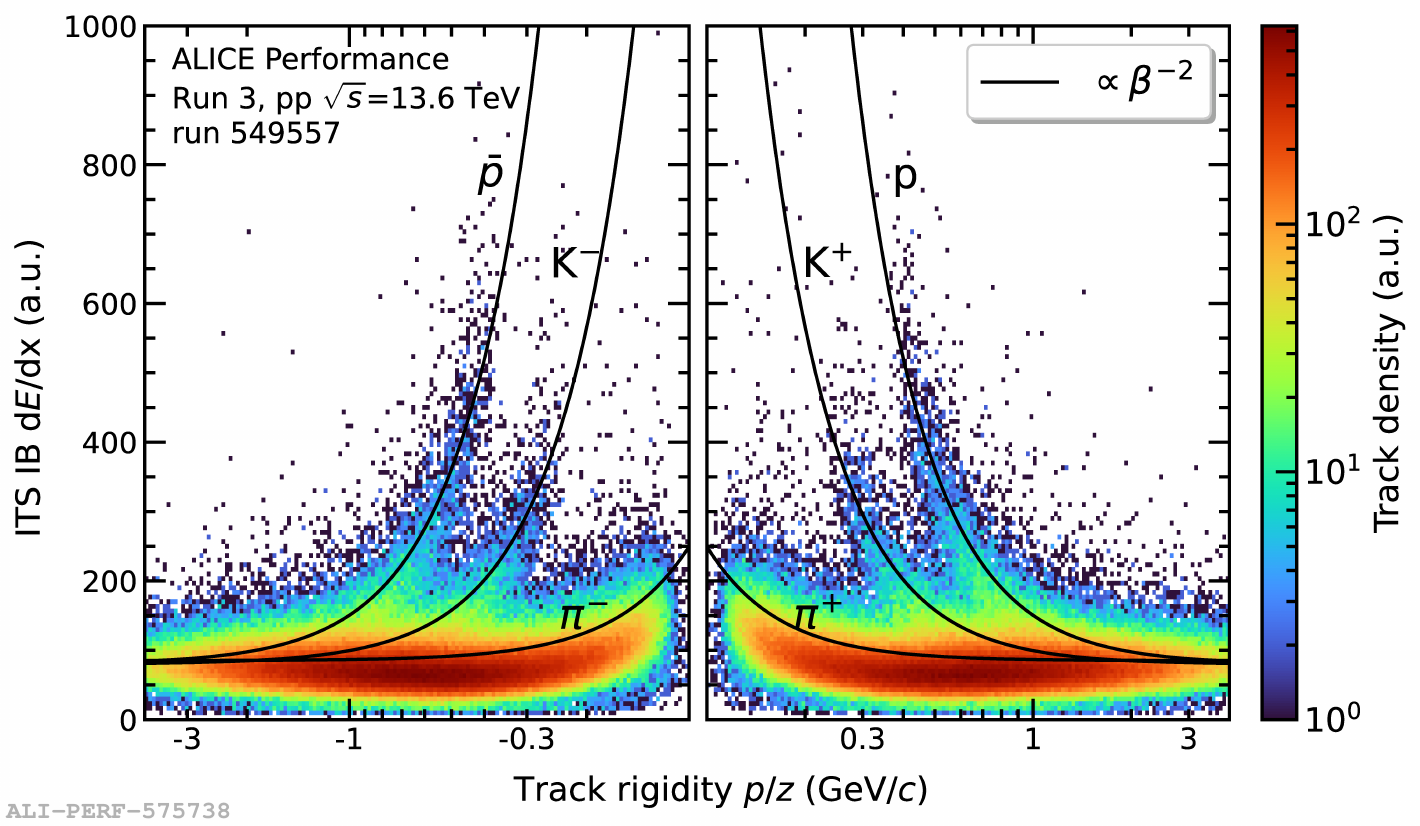}
    \caption{\dedx spectrum versus track rigidity in the ALICE ITS2 Inner Barrel. The lines are a parametrization of the detector response with a function proportional to $1/\beta^2$. }
    \label{fig:dedx}
\end{figure}

The ALPIDE chip features a digital binary readout, without the possibility to measure the Time over Threshold (ToT) and to perform Particle IDentification (PID) during standard data taking and with standard chip settings. In this Section, a method using a dedicated setting of the ALPIDE allowing us to measure the TOT and to convert it into a charge to obtain PID information exploiting the ITS2 IB, is shown. This method only works at low occupancy and is therefore not available during normal ALICE operation. To convert the ToT in a charge, it is essential to obtain the full signal shape. For normal operation, the analogue response of the ALPIDE is clipped to a maximum length of about 5 $\upmu$s. In order to obtain a pulse length corresponding to the collected charge, the clipping is deactivated. 
This way, the signal will be present in multiple subsequent readout frames, and then it is possible to count for how long the pixel stays on by oversampling the signal. This is the only way to measure the ToT given the binary readout of the pixels.
To achieve higher ToT resolution, the signal must be oversampled at a high framing rate. For this study, a run performed at a 2.2~MHz read-out rate was used, which is about ten times the standard 202 kHz at which data are read during normal operations. 
However, oversampling increases the data volume significantly, which poses the risk of filling all three event buffers in each pixel, leading to a "busy" condition where the chip cannot accept other triggers, leading to data loss. To mitigate this issue, the data-taking for the \dedx measurement was performed with the lowest interaction rate reachable by LHC in pp collisions, of about 1 kHz, 500 times lower than the standard interaction rate in ALICE of 500 kHz. 
\\
By measuring the signal shape while injecting known charges into the pixels it has been found that the ToT is proportional to the charge deposited on the pixel, then after a pixel-by-pixel calibration, it is possible to obtain the charge deposited on the single pixels. Since the charge released by each particle track traversing the sensor is shared among adjacent pixels, it is possible to calculate the total charge released after the clusterization of the pixels, summing the charge of the single pixels belonging to the same cluster.
\\
The charge released is proportional to the specific energy loss per unit length $dE/dx$. For each particle track traversing the detector, the \dedx has been calculated through the truncated mean of the charge associated with each cluster in each of the three layers of the IB. This method was used to cut out the Landau tails characterizing the \dedx spectrum, in order to improve the PID resolution. The resulting charge was then corrected for the angle of the track. Figure \ref{fig:dedx} shows the \dedx spectrum obtained with ALPIDE MAPS using this method.
The lines are a parametrization of the detector response with a function proportional to $1/\beta^2$:

\begin{equation}
    \frac{dE}{dx} \propto \frac{1}{\beta^2} \quad \quad \mathrm{with} \quad \quad \beta = \frac{v}{c}=\frac{p}{E}=\frac{p}{\sqrt{m^2+p^2}}
\end{equation}
Taking account of the respective masses of pions, kaons and protons, which feature the spectrum of particles measured in ALICE, it was possible to confirm that the three bands characterizing the plot matched the typical spectrum of the particles measured in the ALICE experiment \cite{PKPi}, and then it was possible to prove that it is possible to use the ALPIDE MAPS to perform PID in specific conditions.

\section{Conclusions}
The ALICE Inner Tracking System was completely replaced during LHC Long Shutdown 2 with a 7-layers pixel-only detector entirely composed of ALPIDE MAPS. This upgrade led to a general improvement of the performance of the ITS, in particular concerning the spatial resolution, impact parameter resolution, and read-out capabilities. The new detector is closer to the interaction point than the previous one, leading to new possibilities of studies involving non-prompt cascades, hypernuclei, and exotic bound states with unprecedented resolution.
\\
To ensure stable operations and high data quality, a regular calibration of the ITS2 has to be performed. The standard calibration procedure involves a threshold calibration, that is composed by a threshold tuning to actually tune the threshold and a threshold scan to monitor it, and a noise calibration to mask the noisy pixels and reduce the detector fake-hit rate. The calibration is a very challenging procedure for the high number of channels involved, and for this reason the calibration scans are performed only on 1-2\% of pixels of the whole detector. Results show a good stability of noise and threshold over time, and thanks to a continuous monitoring of these parameters it is possible to retune the detector if the threshold and noise levels deviate from the standard ones. In particular, the threshold is maintained around 100 \el with periodic threshold tuning around once per year, and the fake-hit rate is maintained below the design value of 10$^{-6}$ hits/event/pixel, with a very low number of masked pixels, through the noise calibration.
\\
The capability of the ITS2 IB to perform particle identification has been demonstrated, being able to discriminate the pions, kaons, and protons spectrum typical of ALICE.


\end{document}